\documentclass[fleqn,usenatbib]{mnras}

\usepackage{newtxtext,newtxmath}
\usepackage[T1]{fontenc}

\DeclareRobustCommand{\VAN}[3]{#2}
\let\VANthebibliography\thebibliography
\def\thebibliography{\DeclareRobustCommand{\VAN}[3]{##3}\VANthebibliography}

\usepackage{graphicx}
\usepackage{amsmath}

\title[Most open clusters are on the RAR and BTFR]{Most open clusters follow the radial acceleration relation (RAR) and the baryonic Tully-Fisher relation (BTFR)}

\author[Huisjes \& Hernandez]{
Mark D. Huisjes,$^{1}$\thanks{E-mail: m.huisjes@degoudsewaarden.nl}
X. Hernandez$^{2}$
\\
$^{1}$CSG De Goudse Waarden Lyceum, Heemskerkstraat 105, 2805SN Gouda\\
$^{2}$Universidad Nacional Autónoma de México, Instituto de Astronomía, A. P. 70-264, 04510, CDMX, México\\
}

\date{}

\pubyear{\the\year{}}

\begin{document}
\label{firstpage}
\pagerange{\pageref{firstpage}--\pageref{lastpage}}
\maketitle

\begin{abstract}
 We test whether parsec-scale stellar systems in the Milky Way follow the galactic radial acceleration relation (RAR) or the baryonic Tully–Fisher relation (BTFR). We analyse 5646 Gaia DR3 open clusters from the Hunt \& Reffert catalogue. Observed accelerations are derived from velocity dispersions and characteristic radii, and baryonic accelerations from stellar masses and characterisitc radii. The clusters are placed on the RAR and BTFR planes and compared with Newtonian and MOND expectations. Approximately 90 per cent of open clusters (those with $N_\star \leq 250$) lie close to the RAR, albeit with significant scatter. In a first-of-its-kind test, a smaller fiducial sample is consistent with a best-fitting acceleration scale $g_\dagger \approx 1.2 \times 10^{-10}\ \mathrm{m\,s^{-2}} \pm 0.5$ dex, compatible with canonical MOND values. More massive clusters approach the Newtonian virial expectation. No correlations are found between RAR residuals and galactocentric radii, distance to the Galactic disk midplane, age, or morphology. Tidal effects and unresolved binaries are insufficient to reproduce the observations without fine-tuning. Interpreted within a MOND framework, the alignment of most open clusters with the RAR and BTFR suggests that low-acceleration dynamics operate on parsec scales within the Milky Way. This implies that the Galactic gravitational field is not smooth on these scales and may include regions where the total gravitational acceleration falls below $a_0$, partially mitigating the external field effect, thereby motivating higher-resolution modelling of the Galactic potential and informing other small-scale gravity tests within the Galaxy.
\end{abstract}




\section{Introduction}
Empirical scaling relations have played a central role in shaping our understanding of galactic dynamics. Two of the most striking are the baryonic Tully–Fisher relation (BTFR),~\citet{2000ApJ...533L..99M}, and the radial acceleration relation (RAR),~\citet{2016PhRvL.117t1101M}. The RAR links the observed gravitational acceleration in galaxies to that predicted from their baryonic mass distribution, while the BTFR connects total baryonic mass to asymptotic rotation velocity through a tight power-law relation. These relations hold across many decades in mass and surface density, with remarkably small intrinsic scatter. They are commonly interpreted either as emergent regularities of galaxy formation within dark matter halos or as manifestations of a modified law of gravity operative at low accelerations.

To date, these relations have primarily been tested on kiloparsec scales in rotating disc galaxies, dwarf spheroidals, and galaxy groups. Much less attention has been paid to whether analogous behaviour persists on parsec scales within galaxies. Open clusters (OCs) provide an opportunity to probe this regime. OCs are self-gravitating systems with characteristic internal accelerations that fall below Milgrom’s constant, $a_0 \sim 1.2\times10^{-10}m\,s^{-2}$, \citet{1983ApJ...270..365M}. Unlike galaxies, however, they reside deep within the gravitational potential of the Milky Way. Their internal dynamics are therefore not only determined by their own mass distributions but also by the external Galactic field in which they are embedded.

Under modified Newtonian dynamics (MOND), this environmental dependence is formalized through the external field effect (EFE). Because the MOND field equation is non-linear, the internal dynamics of a system depend on the total gravitational field, including any ambient external contribution. A system whose internal accelerations are below $a_0$ can nevertheless behave in a (quasi-)Newtonian manner if it is embedded in an external field comparable to or exceeding $a_0$. In the context of the Milky Way disc, the typical large-scale gravitational field near the solar circle is $\sim1.5a_0$. Under the standard AQUAL and QUMOND Lagrangian formulations of MOND, open clusters located in such an environment are therefore expected to exhibit no MOND effects.

This theoretical expectation leads to a sharp test. If open clusters in the Galactic disc systematically follow the deep-MOND branch of the RAR despite being embedded in an external field of order $a_0$, then either the EFE operates differently than currently formulated, or the level of small-scale density fluctuations present in the galactic disc (e.g. \citet{Poggio21}, \citet{Ge24}) lead to regions where the total gravitational potential is much smaller than $a_{0}$.

Conversely, if their dynamics are fully consistent with Newtonian expectations once tidal heating and the kinematic effects of unresolved binaries on the observed velocity dispersion are accounted for, then parsec-scale systems do not provide additional constraints on theories of gravity.

The central question addressed in this paper is therefore whether open clusters, when placed in the acceleration plane defined by the RAR and in the mass–velocity plane defined by the BTFR, align or not with the same relations observed in galaxies, and what this implies for the structure of the Galactic gravitational field. We analyse a large, homogeneous sample of 5646 open clusters from \cite{2021A&A...646A.104H, 2023A&A...673A.114H, 2024A&A...686A..42H} to determine their observed kinematic accelerations and baryonic gravitational accelerations using consistent definitions. By comparing their locations in the RAR and BTFR planes to both Newtonian and MOND expectations, we test whether parsec-scale systems behave as if they are embedded in a smooth external field or instead generally occupy regions where the effective acceleration drops below $a_0$.

This approach allows us to analyse the relation between internal dynamics, external gravitational environment, and empirical scaling relations that were originally established on galactic scales.

\section{MOND theory and the external field effect}
Modified Newtonian Dynamics (MOND) is a modification of classical gravity proposed by Mordehai Milgrom in 1983 to account for the observed kinematics of galaxies without invoking non-baryonic dark matter \citep{1983ApJ...270..365M, 1983ApJ...270..371M, 1983ApJ...270..384M}. The central empirical motivation is the persistence of approximately flat rotation curves in disc galaxies at large radii, where the Newtonian gravitational acceleration inferred from the observed baryonic mass distribution would predict a declining orbital velocity. MOND posits that the law of gravity is modified only in regimes where the characteristic acceleration falls below a universal constant $a_{0} \approx 1.2 \times 10^{-10}\,\mathrm{m\,s^{-2}}$, while standard Newtonian gravity is recovered at accelerations well above this scale.

In MOND, the gravitational acceleration $g_M$ experienced by a test particle deviates from the Newtonian value $g_N$ when $g_N \lesssim a_0$. The modification is commonly written in implicit form as
\begin{equation}
\mu\!\left(\frac{g_M}{a_0}\right) g_M = g_N
\end{equation}
where $\mu(x)$ is an interpolating function satisfying $\mu(x) \to 1$ for $x \gg 1$ (recovering Newtonian gravity) and $\mu(x) \to x$ for $x \ll 1$ (the deep-MOND regime). In the limit $g_N \ll a_0$, this yields
\begin{equation}
g_M \approx \sqrt{g_N a_0}
\end{equation}
Equivalently, one may express the relation for convenience using the inverse interpolation function $\nu(y)$, defined such that
\begin{equation}
\label{nu-eq}
g_M = \nu\!\left(\frac{g_N}{a_0}\right) g_N
\end{equation}
with $\nu(y) \to 1$ for $y \gg 1$ and $\nu(y) \to y^{-1/2}$ for $y \ll 1$. This formulation emphasizes that MOND modifies the effective gravitational field at low accelerations rather than introducing additional unseen mass.

A principal consequence of the deep-MOND scaling $g_M \propto \sqrt{g_N}$ is that the gravitational field around an isolated mass declines asymptotically as $1/r$ rather than $1/r^2$. For a point mass $M$, the asymptotic circular velocity becomes radius-independent and satisfies
\begin{equation}
V^4 = G M a_0
\end{equation}
which directly yields the baryonic Tully--Fisher relation with Milgrom's constant as the normalization parameter. More generally, MOND permits rotation curves to be computed from the observed baryonic mass distribution with essentially one additional constant, $a_0$. Conversely, observed rotation curves can be inverted to infer baryonic surface density profiles. On galactic scales, this predictive capacity is a defining empirical feature of the theory. 

At galaxy cluster scales, however, MOND has typically been seen to fall short in fully explaining observed dynamics by still requiring a factor of 2 more mass than present in the baryons, although recently  \citet{Zhang26} have shown that for IMF choices restricted by the observed metallicities of the cluster galaxies, enough stellar remnants might be expected in the intra-cluster medium to significantly ameliorate this issue. Further, as a non-relativistic theory, MOND is unsuitable for addressing cosmological or gravitational lensing observations in general. 

MOND can be formulated as a modified gravity theory derived from a Lagrangian. The original non-relativistic formulation, known as AQUAL (AQUAdratic Lagrangian) by \citet{1984ApJ...286....7B}, replaces the Newtonian gravitational action with a non-linear functional of the gradient of the potential:
\begin{equation}
\mathcal{L} \propto - a_0^2\, F\!\left(\frac{|\nabla \Phi_{tot}|^2}{a_0^2}\right)
\end{equation}
where the function $F$ is chosen so that the resulting field equation below reproduces the $\mu$-interpolation relation. This yields a modified Poisson equation of the form
\begin{equation}
\nabla \cdot \left[ \mu\!\left(\frac{|\nabla \Phi_{tot}|}{a_0}\right) \nabla \Phi_{tot} \right] = 4\pi G \rho
\end{equation}
An alternative but dynamically equivalent formulation by \citet{2010MNRAS.403..886M}, QUMOND (quasi-linear MOND), derives from a Lagrangian constructed so that the non-linear aspects are shifted into an algebraic relation involving the Newtonian potential. In QUMOND, one first computes the Newtonian potential from the baryonic mass distribution and then applies the $\nu$-function to obtain the MOND correction, leading to a linear Poisson equation with an effective source term. Both AQUAL and QUMOND are non-relativistic Lagrangian realizations of MOND and ensure the usual conservation of energy and momentum within the modified gravitational framework.

While MOND has occasionally been explored as a modification of inertia, e.g. \citet{Milgrom22}, in which the force--acceleration relation itself is altered, this interpretation remains less developed. The standard treatment regards MOND primarily as a non-linear modification of the gravitational field operative below a universal acceleration scale.

\subsection{The external field effect}
\begin{figure}
   \centering
   \includegraphics[width=\hsize]{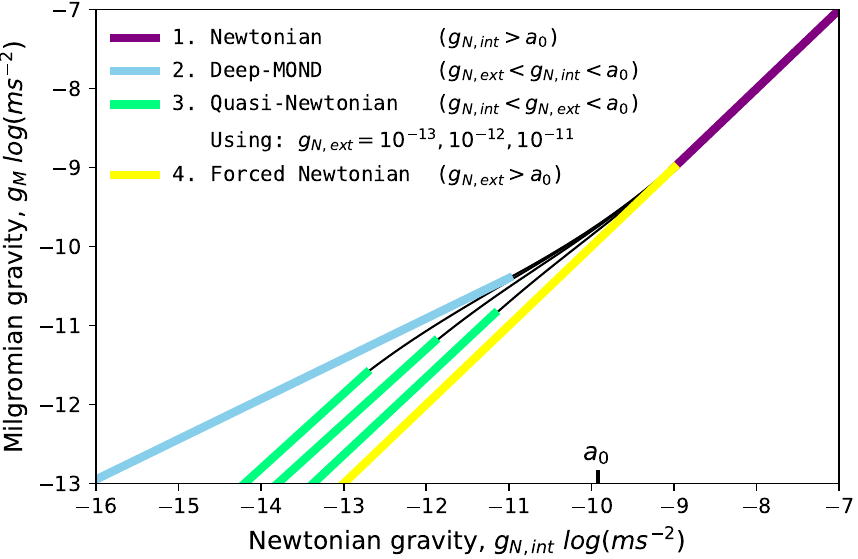}
      \caption{The four regimes of MOND, based on figure one of \citet{2022Symm...14.1331B}.}
         \label{Fig_FourRegimes}
   \end{figure}

Because the MOND field equation is non-linear and depends on the \textit{total} gravitational field, the internal dynamics of a self-gravitating system depend not only on its own mass distribution but also on any ambient gravitational field in which it is embedded. This phenomenon, known as the external field effect (EFE), arises even if the external field is spatially uniform and tidal forces are absent. Such behaviour has no equivalent in Newtonian gravity. A system whose internal accelerations are below $a_0$ may nonetheless behave in a Newtonian manner if it resides in an external field exceeding $a_0$. The external field effectively suppresses the deep-MOND enhancement. When $g_{int}<g_{ext}<a_0$ the Milgromian enhancement is only partially suppressed and causes the internal dynamics to interpolate between Newtonian and Milgromian behaviour according to the total gravitational environment. This is called the quasi-Newtonian regime and is effectively equivalent to rescaling the gravitational constant G. For a graphical overview of the possible regimes see Fig. \ref{Fig_FourRegimes}.

The EFE reflects a violation of the strong equivalence principle while preserving the weak equivalence principle\footnote{For a discussion on the various formulations of the equivalence principle see \citet{2015AmJPh..83...39D}.}, as all test bodies still follow identical trajectories in a given gravitational field. The EFE has concrete implications for open clusters in galactic discs, dwarf satellite galaxies within host systems, and wide binary stars in the Milky Way, where the background galactic field can influence internal kinematics.

The EFE can be approximated to first order for a system consisting of an internal and external Newtonian gravitational field contribution by changing equation \ref{nu-eq} to the form below based on equation 60 by \citet{2012LRR....15...10F}.
\begin{equation}
\label{efe-eq}
g_{\mathrm{M}} =
\Bigl(g_{\mathrm{N,int}} + g_{\mathrm{N,ext}}\Bigr)\,
\nu\!\left(\frac{g_{\mathrm{N,int}} + g_{\mathrm{N,ext}}}{a_{0}}\right)
- g_{\mathrm{N,ext}}\,
\nu\!\left(\frac{g_{\mathrm{N,ext}}}{a_{0}}\right)
\end{equation}

There is now considerable observational evidence in favor of the EFE under MOND interpretations of astronomical data. For example observations by \citet{2013ApJ...766...22M, 2013ApJ...775..139M} support the EFE in the velocity dispersions of dwarf satellite galaxies embedded in the gravitational field of their host. The EFE has also been detected in the outer edges of spiral galaxies by \citet{2020ApJ...904...51C} where rotation curves are affected by the gravitational field of the large scale structure of the universe. Recently, studies of the anomalous relative velocity between components of wide binaries in the Solar neighbourhood have yielded results consistent with the presence of standard MOND EFE phenomenology e.g.
\citet{Hernandez24}, \citet{Chae24b}. Finally laboratory Cavendish experiments using a torsion pendulum also point to the absence of any enhanced motion at low internal accelerations indicating either ordinary Newtonian mechanics or the presence of the external field effect due to the very strong gravitational field of the Earth \citep{2007PhRvL..98o0801G}.

Although open clusters were already mentioned in the foundational papers on MOND \citep{1983ApJ...270..365M}, analyses of these systems has remained sparse until recently. Open clusters have long been considered nearly Newtonian because of relatively small excess velocity dispersions and the dominant influence of the Galactic gravitational field. Indeed, open clusters were the original motivation for including the EFE in the initial development of MOND \citep{1983ApJ...270..365M}. According to Milgrom, Salpeter and Tremaine pointed out that open clusters in the solar neighbourhood showed no appreciable mismatch between their kinematic and stellar masses, despite having internal baryonic accelerations well below $a_0$. Of course, it is well known that under a Newtonian interpretation, open clusters are in the process of dissolving into the field, and indeed, abundant evidence exists showing these systems not to be equilibrium bound self-gravitating objects often presenting clear tidal features. It remains to be seen to what degree their present internal kinematics might retain information of their formation processes, a point which we investigate in detail in what follows.

\section{Sample selection} \label{sample}
We begin with the star cluster catalogue compiled by \citet{2021A&A...646A.104H, 2023A&A...673A.114H, 2024A&A...686A..42H}, henceforth HR134. This enables a systematic investigation of the dynamics of stellar subsystems in the Milky Way. The catalogue of clusters created by HR134 contains several types of over-densities such as what HR134 classifies as open clusters, moving groups, globular clusters and a "rejected" category that was classified as such manually. In this study we will only consider the 5646 clusters HR134 classifies as open clusters.

\subsection{Data quality cuts}\label{selection}
Two quality cuts were applied to the HR134 open clusters. The first is a distance cut, eliminating all clusters with a heliocentric distance greater than $3\ $kpc because beyond this distance completeness decreases and faint magnitude systematics become important contributors to the measured proper motion. This removes 1573 clusters from the sample. The second cut is to remove all clusters more than $150\ $pc away from the midplane of the disc as these are likely on orbits which cross the disc periodically and are particularly prone to tidal disruption. This removes another 454 clusters from the sample leaving a total of 3618 open clusters to be analysed.

The sample of 3618 open clusters is divided into three parts for analysis based on the number of resolved stars $N\star$. The first group has $N\star \leq250$ and comprises 90\% of the full science sample. The second group has $250<N\star\leq500$ and accounts for another 6\%. The third and final group has $N\star>500$ and consists of the remaining 4\% of clusters.

A further fiducial high quality sample was selected from the $N\star\leq250$ group which contained 2423 open clusters by requiring that each have a colour-magnitude diagram quality index exceeding 75\%. This index, used by HR134, compares the CMD of an observed cluster to the best-fitting isochrone for it and estimates the fraction of stars which are consistent with this optimal isochrone, given observational errors, and the fraction which are likely foreground or background field contaminant stars. A value of 100\% denotes all stars being likely cluster members consistent with the optimal isochrone, while a value of 0 would denote a cluster where all assigned member stars are probably contaminant non-cluster members.

\begin{figure*}
   \centering
   \includegraphics[width=\hsize]{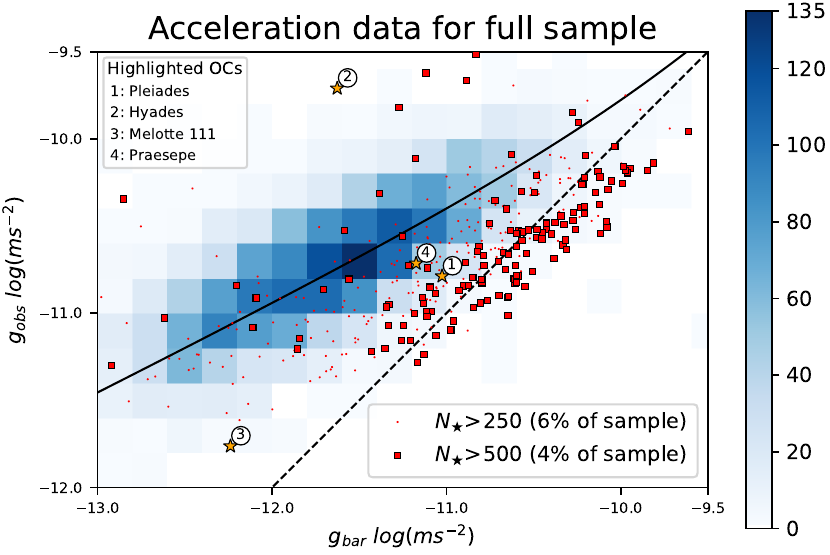}
      \caption{Comparison of observed kinetic accelerations, $g_{\mathrm{obs}}$, and Newtonian virial equilibrium accelerations, $g_{\mathrm{bar}}$ for the inferred masses, velocity dispersions and radii for the 3618 open clusters after our quality cuts from the HR134 catalogue. The colour number density shading gives the position of the 3251 $N_{\star}\leq 250$ clusters which accurately trace the galactic RAR relation given by the solid curve. The solid squares show the 145 most massive $N_{\star}\geq 500$ clusters, mostly tracing the Newtonian virial expectations, while the dots give the positions of the 222 intermediate mass $250 \leq N_{\star} \leq 500$ clusters, filling the region in between the two previous populations.}
         \label{Fig_Main_plot}
   \end{figure*}

\subsection{Derived quantities}
The heliocentric distance to each open cluster is estimated independently using two methods: a photometric method based on the distance modulus, and a geometric method based on parallax. An accurate distance is necessary to convert the measured proper motions to physical velocities.

For the photometric estimate, the 16th, 50th, and 84th percentiles of the distance modulus distribution given by the HR134 catalogue are converted into distances via
\begin{equation}
d = 10^{(m - M)/5 + 1} \, ,
\end{equation}
yielding three corresponding distance estimates. A symmetric uncertainty is then defined as the mean of the upper and lower absolute deviations from the median distance. An analogous procedure is applied to the parallax-based distance estimate, again defining a central value and a symmetric uncertainty from its percentile range.

The two distance estimates are then combined using inverse-variance weighting. Defining the weights as 
\( w_{\mathrm{phot}} = 1/\sigma_{d,\mathrm{phot}}^{2} \) 
and 
\( w_{\pi} = 1/\sigma_{d,\pi}^{2} \),
the joint distance with standard error propagation is given by
\begin{equation}
d_{\mathrm{joint}} =
\frac{d_{\mathrm{phot}}\, w_{\mathrm{phot}} + d_{\pi}\, w_{\pi}}
     {w_{\mathrm{phot}} + w_{\pi}} \, ,
\end{equation}
\begin{equation}
\sigma_{d,\mathrm{joint}} =
\sqrt{\frac{1}{w_{\mathrm{phot}} + w_{\pi}}} \, .
\end{equation}

For the sample considered in this work, the photometric and parallax distances are nearly identical for most open clusters. Only beyond heliocentric distances of approximately $3\,$ kpc do the two methods begin to diverge at the order-of-unity level. As a result, the joint estimator typically differs only marginally from either individual estimate within the bulk of the sample. As in HR134, raw Gaia parallaxes were corrected for the parallax zero point bias described by \citet{2021A&A...649A...2L} when calculating parallax distances. 

The velocity dispersion given by the HR134 catalogue is calculated by taking the standard deviation of the proper motions. According to Section 3.3 of \citet{2023A&A...673A.114H} mean cluster parameters were calculated over only those stars assigned to the cluster that have membership probabilities larger than 50\% and are within the tidal radius of the cluster, so as to exclude any tidally perturbed stars within the tidal tails. This tidal radius is defined in a model independent way as the radius at which the overall cluster has the best contrast to field stars using the CST parameter produced by the HDBSCAN clustering algorithm. According to HR134, this is similar to the King (1962) definition of tidal radius as the radius at which a cluster’s density begins to exceed that of the density of the field. This was verified by recalculating the velocity dispersion from the individual member star proper motions using the "inrt" and "inrj" parameters of the catalogue which classify a star as being in- or outside of the tidal or Jacobi radius respectively. Further excluded are all those stars with $RUWE>1.4$. The catalogue does not weight the proper motions by their uncertainties in calculating the velocity dispersion. Here we adopt the velocity dispersion given by HR134 for each cluster, $\sigma_{\star}$.

We further recalculated $\sigma_{\star}$ for each cluster directly from the individual member stars to test its sensitivity to using different selection criteria and methods of calculation. To test how sensitive the velocity dispersion is to the HDBSCAN derived membership probability, a magnitude cut $12>G>17$ was used instead, along with excluding all stars outside the tidal radius and with a $RUWE>1.4$ and any with an available non-single star solution in Gaia DR3. This alternative selection did not significantly alter the velocity dispersion inferred for the clusters. A further cut excluding stars with $RUWE>1.2$ only lowered the velocity dispersion by about 10\%.

Following HR134, these recalculations corrected the raw Gaia proper motions for the proper motion bias of bright Gaia sources identified by \citet{2021A&A...649A.124C}. A further correction for the secular aberration drift due to the solar system's acceleration towards $SgrA*$ identified by \citet{2021A&A...649A...9G} was also tested which did not noticeably alter the velocity dispersion when applied or not.

For each open cluster the baryonic gravitational and observed kinematic acceleration were determined based on the definitions of \citet{2017ApJ...836..152L}:
\begin{equation}
g_{\mathrm{obs}} = -\nabla\Phi_{tot}(R_{\mathrm{1/2}}) = \frac{\left(\sqrt{3}\sigma_{\star}\right)^{2}}{R_{\mathrm{1/2}}}
\end{equation}
Where $\sigma_{\star}$ is the mean one dimensional velocity dispersion of the stars, $R_{\mathrm{1/2}}$ is the half member radius and the factor $\sqrt{3}$ converts the one dimensional velocity dispersion to the full 3D velocity dispersion assuming isotropy. Instead of the factor $\sqrt{3}$, \citet{2021AJ....162..202M} suggests to use 2 as this empirically shifts dwarf galaxies onto the baryonic Tully-Fisher relation. He also notes that MOND predicts a factor of 2.12 for this parameter for deep-MOND isolated systems. Here we adhere to Lelli's definition through the assumption of isotropy, in order to use a strict kinematic acceleration measure without introducing any relations like the RAR or the BTFR or any assumptions on the possible validity of MOND a priori. To convert the proper motions to physical units the joint distance mentioned previously was used.

The baryonic gravitational acceleration is computed from the total cluster mass $M$ (in solar masses), assuming spherical symmetry and evaluating the Newtonian acceleration at the same characteristic radius as used above. Following \citet{2017ApJ...836..152L}, we adopt the factor 2 in the denominator appropriate for using the half-mass radius proxy:
\begin{equation}
g_{\mathrm{bar}} = -\nabla\Phi_{bar}(R_{\mathrm{1/2}}) = \frac{G\,M}{2R_{\mathrm{1/2}}^{2}}
\end{equation}

Hence, as described in this section, we obtain a set of ($\sigma_{star}$, $M$, $R_{1/2}$, $g_{\mathrm{obs}}$, $g_{\mathrm{bar}}$) parameters for each of the 3618 open clusters clearing the quality cuts imposed on the HR134 catalogue. 

\section{Results}
We begin with Fig. \ref{Fig_Main_plot} which shows the inferred $g_{\mathrm{obs}}$ and $g_{\mathrm{bar}}$ values for the three populations described previously, selected through a ranking on the number of stars present per cluster. We see the most massive ones well described by Newtonian virial equilibrium expectations, with the addition of a small distribution towards super-virial values in $g_{\mathrm{obs}}$ reflecting higher $\sigma_{\star}$ values than what corresponds to the dashed Newtonian line shown. At the other extreme, the smallest clusters present a distribution following the radial acceleration relation, an empirical scaling well established at galactic scales, and clearly tracing MOND predictions of four decades ago. This RAR was originally found at baryonic mass scales some eight orders of magnitude above the small OCs studied here. In between the previous two populations we find intermediate mass clusters which populate the region between the two previously described, with a small dispersion extending towards either of the other two regions, as seen in Fig. ~\ref{Fig_Main_plot}

This division of the sample into three groups was done because they show markedly different positions in the acceleration plane. The exact boundaries between these groups are somewhat arbitrary as the characteristics of these clusters change smoothly up towards the highest number of resolved stars (and highest mass). While the specific membership thresholds chosen are arbitrary, the three distinct populations shown are robust to small changes in the details of these thresholds.

Although it might be more physically motivated to divide the sample based on the inferred stellar mass of the cluster, this would add numerous inferential steps each with its own assumptions and uncertainties. In practice, it was found that using the observational quantity $N\star$ provides a cleaner separation in the acceleration plane.

   \begin{figure}
   \centering
   \includegraphics[width=\hsize]{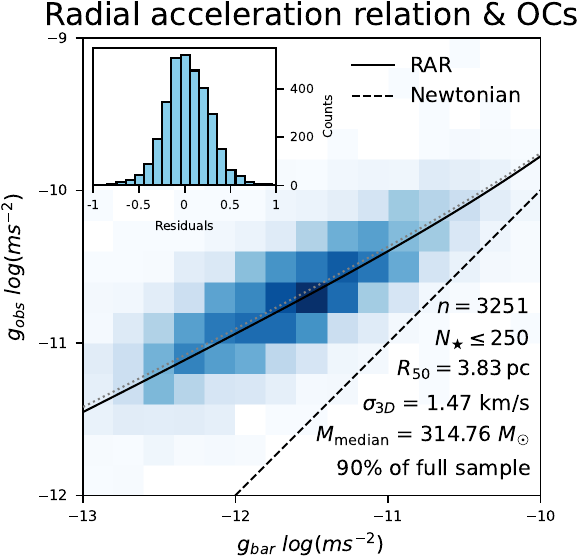}
      \caption{Same data as shown in Fig.  \ref{Fig_Main_plot} but only for the smallest $N_{\star} \leq 250$ clusters. The thin dotted line gives the resulting fit to these clusters to eq.~\ref{rar_eq}, yielding an optimal fit acceleration scale almost identical to the standard $a_{0}$ value of MOND, which in turn results in the galactic RAR given by the solid curve. The equivalent fit using the fiducial $N_{\star}<250$ sub-sample with CMD quality$>75\%$ clusters actually coincides with this solid curve.}
         \label{Fig_OC_RAR}
   \end{figure}

   \begin{figure}
   \centering
   \includegraphics[width=\hsize]{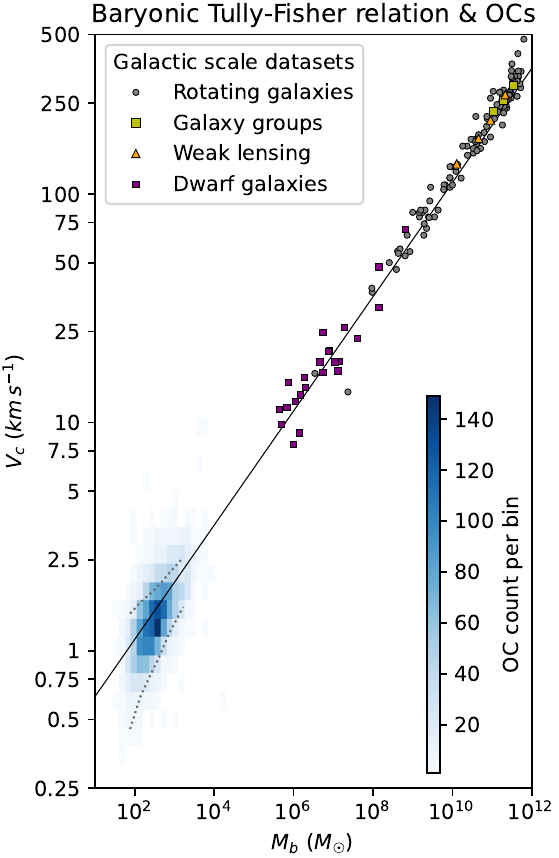}
      \caption{Open clusters with $N\star<250$ compared to the baryonic Tully-Fisher relation. The error bars indicate the $1\sigma$ spread of the data for each mass bin. The dotted lines are the systematic uncertainty expected due to the presence of unresolved binary contamination, as modelled by HR134. Other data included: galaxy groups~\citep{2019PhRvD..99d4041M, 2022A&A...662A..57M}, rotating galaxies~\citep{2021MNRAS.507.5820D, 2019MNRAS.484.3267L, 2021AJ....162..202M}; dwarf galaxies~\citep{2021AJ....162..202M}; weak lensing \citep{2024ApJ...969L...3M}.}
         \label{Fig_BTFR}
   \end{figure}

We now turn our attention to a more careful exploration of the $N_{\star} \leq 250$ sample, shown in Fig.  \ref{Fig_OC_RAR}. This figure presents a zoom of the previous figure, together with, for the first time in the literature, a fit using this sample to the radial acceleration relation leaving $g_\dagger$ as a free parameter,

\begin{equation}
\label{rar_eq}
g_{\rm obs} = \frac{g_{\rm bar}}{1 - e^{-\sqrt{g_{\rm bar}/g_\dagger}}}.
\end{equation}

The inferred value we obtain is of an acceleration scale $g_\dagger=1.4\times10^{-10} m\,s^{-2} \pm0.5dex$. 
Turning to the fiducial $N_{\star} \leq 250$ sub-sample with a cleaned CMD the result of this inference becomes $g_\dagger=1.2\times10^{-10} m\,s^{-2} \pm0.5dex$. 
This matches the canonical MOND value for this quantity as first measured by \citet{1991MNRAS.249..523B}, namely $a_0=1.21\times10^{-10} m\,s^{-2}$ but vastly less precise than modern values such as $a_0=1.20\pm0.02\times10^{-10} m\,s^{-2}$ by \citet{2017ApJ...836..152L}. In Fig.  \ref{Fig_OC_RAR} the first fit is shown by the small dotted line, while the second is indistinguishable from the standard galactic RAR (again shown by the solid curve). The inset in Fig. ~\ref{Fig_OC_RAR} gives the distribution of residuals from the best fit RAR relation, which closely follow a Gaussian distribution, showing the data to be consistent with the fit presented to within effective errors in the determination of the plotted quantities.

It is impressive that taking a sample of Galactic open clusters with a typical mass of a couple hundred stars, this fit should yield a central inferred acceleration scale almost indistinguishable from what results when describing the rotation curves of spiral galaxies.

Comparing Fig. \ref{Fig_Main_plot} with Fig. \ref{Fig_FourRegimes} under a MOND perspective, we would be forced to conclude that the $N_\star\leq250$ population lies in (or retains memory of a formation phase under) the deep-MOND regime, the $250<N\star\leq500$ population resides within the quasi-Newtonian regime and the $N_\star>500$ population probes the forced Newtonian regime, the latter two of which are expressions of the external field effect.

A final test of the emergent deep-MOND character of our $N_\star\leq250$ population is explored in Fig. ~\ref{Fig_BTFR}, where we plot the mass and geometric equivalent rotation velocities of these clusters, $V_{c}=\sqrt{3} \sigma_{\star}$, large blue squares, in comparison, the corresponding values for galactic systems defining the baryonic Tully-Fisher relation. The dotted lines give 500 m/s error ranges on the data presented, where this value represents the expected uncertainty due to the presence of unresolved binaries. While this uncertainty will only result in a possible reduction of the true cluster velocity dispersion, we include it symmetrically as a generous estimate of the systematic error budget of the problem. Under a MOND perspective it has been shown (e.g. \citet{2021AJ....162..202M}), that a conversion factor between line-of-sight velocity dispersion and equivalent BTFR circular velocity of 2.12 rather than $\sqrt{3}$ is better physically motivated, using this slightly larger factor brings the mean velocity values for the open clusters treated into even better agreement with the galactic BTFR than what is shown in Fig. ~\ref{Fig_BTFR}.

We note also that the masses we are using were derived by HR134 through integrating an assumed IMF suitable to the observed stellar population of these clusters. However, no allowance for mass segregation was included, so that if any relevant stellar mass segregation processes have taken place, a systematic error in mass will be introduced. Of course, any such effect is expected to be relevant only for the densest and oldest clusters treated here, and not for the very small and young $N_\star\leq250$ objects shown in this figure.

\section{Newtonian interpretations of the scaling relations found}
The galactic radial acceleration relation and the baryonic Tully-Fisher relation are predictions of MOND for isolated systems which have been confirmed empirically. Under a Newtonian framework, these are interpreted as the presence of a dominant dark matter halo with properties tightly correlated with those of the observed baryonic galaxies. Since the open clusters examined here clearly follow these relations, one possible interpretation for the observed velocity dispersion excess compared to the Newtonian expectation is that the isolated deep-MOND regime is at work. 

Before interpreting the data presented as evidence for MOND, we must consider that the excess velocity dispersions of open clusters can also arise from several mechanisms unrelated to modified gravity. It is hence necessary to examine whether observational biases, dynamical misclassification, stellar multiplicity, or the presence of a hypothetical dark matter component could plausibly account for the effect within standard gravitational dynamics. The following subsections assess all these possibilities in turn, focusing on whether they can reproduce not only an overall elevation above the Newtonian virial line, but specifically considering the observed alignment with the radial acceleration relation.

   \begin{figure*}
        \centering
        \includegraphics[width=\hsize]{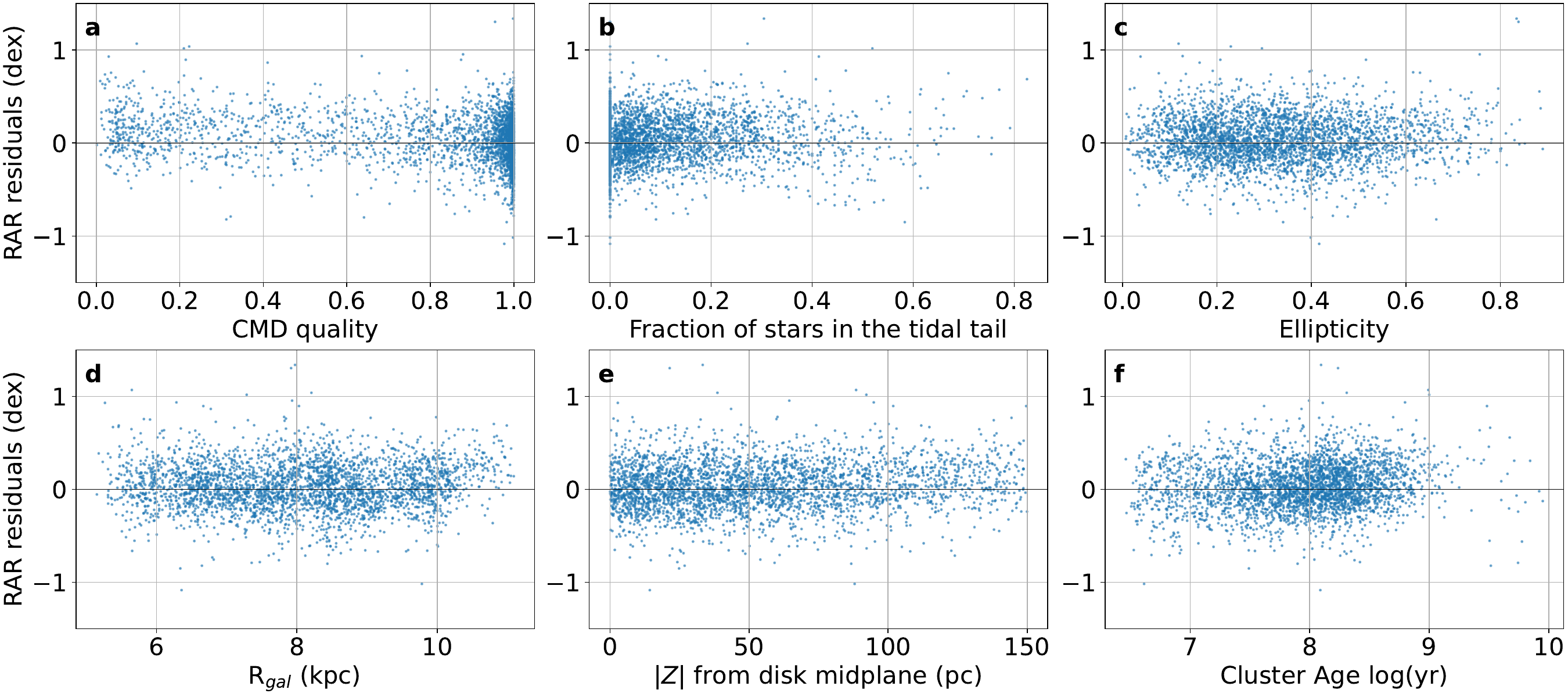}
        \caption{Residuals with respect to the radial acceleration relation for our main  $N_\star\leq250$ open cluster sample, plotted against various independent cluster characteristics.}
        \label{FigScatterPlots}
    \end{figure*}

\subsection{Interloper stars}
Cluster membership in the catalogue of HR134 is determined using the HDBSCAN clustering algorithm applied in position--proper motion--parallax space. Only stars with membership probabilities larger than 50\% are retained for the computation of mean cluster parameters, and only those lying within a model-independent tidal radius defined via the cluster--field contrast (CST parameter) are included. In addition, stars with $\mathrm{RUWE} > 1.4$ are excluded in order to suppress astrometric outliers.

Recomputing the velocity dispersions using alternative selection criteria, including magnitude cuts and stricter $\mathrm{RUWE}$ thresholds, does not significantly change the inferred dispersions. This indicates that contamination by field interlopers is not the dominant driver of the excess velocity dispersion.

If interlopers were responsible for the elevated dispersions, one would expect clusters with cleaner colour--magnitude diagrams to move closer to the Newtonian virial expectation. However, the fiducial high-quality subsample restricted to high CMD quality ($>75\%$) continues to follow the radial acceleration relation. The systematic alignment of low-$N_\star$ clusters with the RAR therefore cannot plausibly be attributed to residual field-star contamination.

This is tested explicitly in Fig. ~\ref{FigScatterPlots} panel a), where we plot the residuals from the best fit RAR as a function of the CMD quality index of the full $N_{\star} <250$ sample. We see that only for the lowest CMD-quality values below 25\% a correlation of the residuals to this quantity appears, for all CMD-quality values above 75\%, residuals to the best fit RAR are distributed symmetrically about the zero line. No minimum CMD quality threshold will remove the clear RAR scaling reported.

\subsection{Unresolved binaries}
Unresolved binary stars with unequal mass ratios shift the system’s centre of light as the components orbit one another. This motion propagates into the measured velocity dispersion of the cluster, inflating it and biasing virial analyses of open clusters if not properly modelled \citep{2010MNRAS.402.1750G}. HR134 attempted to correct for unresolved binaries by adopting mean binary parameters (e.g. period and eccentricity) for field stars from \citet{2017ApJS..230...15M} but found this to be impossible. They warn that the velocity dispersions are easily contaminated and that this can contribute $\gtrsim 500m\,s^{-1}$. This range is included as the dotted lines in Fig.  \ref{Fig_BTFR} and corresponds very well to the observed scatter in the data. It is inconsistent as an explanation for the observed trend.

HR134 notes that, for their sample of open clusters, the simulations used to estimate the influence of unresolved binaries may overestimate the effect. Stars likely to be binaries with poor-quality astrometry were already removed by the \citet{2022MNRAS.510.2597R} selection, in addition to the exclusion of stars with $\mathrm{RUWE} > 1.4$ in the velocity dispersion analysis. Indeed, \citet{Belokurov20} identify this RUWE threshold as a secure limit to exclude unresolved binaries from the Gaia catalogue.

Tightening the RUWE threshold to 1.25 or even 1.0 reduced the inferred cluster velocity dispersions only by about 10\%, and does not alter any of the scalings described in the previous section. Nevertheless, the binary fraction may be higher in denser environments such as open clusters. HR134 concludes that a definitive removal of the impact of unresolved binaries on cluster velocity dispersions may not be possible before Gaia DR4, which is expected to provide substantially more non-single-star solutions than Gaia DR3.

Although unresolved binaries cannot presently be ruled out as a systematic leading to the high virial parameters of open clusters, this interpretation appears unlikely as it would not result in the precise RAR and BTFR scalings we obtain. Open clusters with lower internal baryonic accelerations exhibit higher virial parameters; attributing this trend to unresolved binaries would imply that more diffuse clusters host a larger binary fraction, while denser clusters contain fewer binaries. Attributing the scalings described to unresolved binaries would be contrived and require highly fine-tuned unresolved binary parameters to be adjusted as a function of the number of stars and radii of each open cluster, across the cluster population examined.

\subsection{Dissolution dynamics and tidal heating}

One possible explanation for the fact that the sample of HR134 lies above the Newtonian line of virial equilibrium is that these systems are not bound open clusters but unbound associations, since we know that the origin of disc field stars is precisely open cluster dissolution. Given that the cluster finding algorithm of HR134 does not select over-densities that lack coherent proper motions the sample cannot be comprised of mere chance associations, as clusters are selected on a position-velocity space. However, tidally disrupted formerly bound systems would certainly present a velocity dispersion excess over Newtonian virial expectations. Actually a majority of the open clusters examined do have some tidal features, even if for most these comprise only a small fraction of the stars. In fact, such tidal features themselves offer interesting tests of gravity. \citet{2018A&A...609A..44T, 2022MNRAS.517.3613K} have shown that tidal tail asymmetries in carefully observed open clusters conflict with Newtonian expectations, while being consistent with MOND expectations. If open clusters are heavily tidally disrupted, it would have to be viewed as an uncanny coincidence that the excess velocities found for the small systems land them on the galactic RAR and BTFR, while then only the more sturdy massive clusters with $N\star>500$ would be gravitationally bound structures.

However, contrary to the above explanation, most systems in the sample are rather spherical and have only a small fraction of stars in what HR134 classifies as a tidal tail, so that tidal disruption does not appear to be a sufficient explanation. This is shown in Fig. ~\ref{FigScatterPlots}, panel b), where we plot the residuals from the best fit RAR against the fraction of stars in the tidal tails of the clusters studied. As can be seen, this fraction is smaller than 0.3 for the greater majority of our sample, and for any tidal tail fraction threshold below 0.2, no systematic trend is present, with clusters showing a symmetric residual distribution about zero. Clearly, no tidal tail fraction exclusion criterion will remove the galactic RAR and BTFR scalings we see. Indeed, the distribution of ellipticities of the open clusters defining the galactic scalings mentioned can be seen in Fig. ~\ref{FigScatterPlots}, panel c), and shows most are not highly elongated with ellipticities below 0.5, showing no systematic trend of the RAR residuals with this tidally sensitive shape parameter.

Also, if the excess velocity dispersion were caused by tidal disruption by the smooth galactic gravitational field, this would necessitate such an effect to grow with decreasing galactocentric radius, as the galactic tidal field increases with decreasing galactocentric radius. This is not the case, as can be explicitly seen in panel d) of Fig.~\ref{FigScatterPlots} where the RAR residuals are shown against the galactocentric radii of our open clusters. For any value of this quantity below 10 kpc, residuals show no trend at all and present a symmetric distribution about zero.

Aside from tidal heating from the radial galactic tides, vertical tidal shocking on crossing the midplane of the disc could also be present. This would be more pronounced for the clusters which rise (or dip) most above (or below) the Galactic midplane. Our quality cut at $|Z|<150\ $pc was explicitly chosen to eliminate this source of excess velocity dispersion. Increasing the severity of this cut to $|Z|<100\ $pc or even $|Z|<50\ $pc does not change the fact that $N_\star\leq250$ clusters lie along the RAR. As can be seen in panel e) in Fig. ~\ref{FigScatterPlots}, no systematic correlation in RAR residuals is seen for any $|Z|$ value below $100\ $pc, which contains the overwhelming fraction of our sample. Keeping as close to the Galactic midplane as desired preserves the RAR and BTFR scalings.

Interpreting these systems as unbound associations does not explain why they scatter around the two relations predicted by MOND for systems in which the external field is negligible and followed by galactic systems. In principle, unbound stellar associations could occupy any position above the Newtonian virial equilibrium line (dashed line in Fig.  \ref{Fig_OC_RAR}). It would be a remarkable coincidence if most such associations happened to be observed at precisely the stage in their dissolution at which they acquire the additional dispersion required to lie close to the radial acceleration relation at the time of observation.

Since none of the residuals shown in Fig. \ref{FigScatterPlots} show any systematic deviations from the RAR, it appears that the galactic scalings which we have found to apply to these clusters within their tidal radii, are not driven by tidal heating and/or dissolution dynamics.

\subsection{Dark matter}

Since the scalings of the RAR and the BTFR found to describe the clusters treated (within their tidal radii) are interpreted within a galactic standard gravity outlook as reflecting the presence of dominant dark matter halos, it might be tempting to explore such an explanation for the open clusters studied here.

Historically, open clusters have been regarded as systems in which dark matter is negligible. The seminal review on open cluster formation by \citep{2003ARA&A..41...57L} does not even mention dark matter at all. More recent reviews e.g. \citep{2020SSRv..216...69A, 2020SSRv..216...64K} even explicitly define open clusters as systems containing no dark matter. This partly follows from independent bounds on the local dark matter density within the galactic disc. For example, at the Solar Neighbourhood, it is well established that any local dark matter density above 0.01 $M_{\odot}$pc$^{-3}$ is excluded by the observed vertical distribution and measured vertical velocities of stars, e.g. \citet{Read14} found $0.005<\rho_{DM}<0.015$ and more recently using Gaia, \cite{Bienayme24} derive $\rho_{DM}=0.0128 \pm 0.0008$, both in units of $M_{\odot}$pc$^{-3}$, for the Solar Neighbourhood. This value would increase slightly towards the galactic centre, but decrease away, so that on average, the open clusters we examine would be expected to coexist with the smooth Galactic dark matter halo having a density of close to 0.01 $M_{\odot}$pc$^{-3}$, at most. Given the scales of our systems of below some $10\,$pc, within their tidal radii, where we derive the scalings described, they would be expected to contain at most 10 $M_{\odot}$ of dark matter, which would represent less than some 10\% of the cluster's mass and hence would not explain any velocity boost above a mere 5\%. 

Aside from the hypothetical smooth dark matter halo component, structure formation simulations predict also the presence of small and dense dark matter sub-halos, e.g. \citep{2008MNRAS.391.1685S}, which depending on the assumed detailed physics for this component, could extend down to the mass range of open clusters. However, any such hypothetical sub-halos would form part of the overall Galactic dark halo distribution and hence cross the Galactic disc practically vertically at close to the virial velocity of the halo of close to 160 km s$^{-1}$ 

In fact, such dark matter sub-halos have been envisioned as potential sources of dynamical heating for the Galactic disc e.g. \citet{Benson04}, and would have no chance of serving as seeds of open cluster formation or becoming embedded with the small stellar structures we are examining. We see that the standard explanation within classical gravity for the galactic scalings we have found at open cluster scales, dominant dark matter distributions, has no place at the scales and Galactic disc settings at which we are currently working. Indeed, beyond the continual lack of any dark matter detection, recently \citet{Hernandez25} have shown that dynamical friction constraints across astronomical scales make the dark matter hypothesis in itself rather dubious.

Finally, the option of the OCs studied having internal dynamics which do not reflect an internal equilibrium dynamical state can also be dismissed. Clusters from the $N_\star\leq250$ sample have ages that are much longer than their crossing times. The median age of this sample is 100 million years while the mean crossing time of these clusters are of the order of a million years, see panel f) in Fig. ~\ref{FigScatterPlots}.

\section{MONDian interpretations of the scaling relations found}

As explained in section~2.1, under a MOND description, whenever a system having internal accelerations below $a_{0}$ is immersed in an external acceleration field comparable or larger than $a_{0}$, MOND effects are suppressed and internally the system behaves in a quasi-Newtonian way. Describing the Galaxy as a purely baryonic system including only the Galactic bulge and a smooth Galactic disc matter distribution, this last typically modelled as a double exponential disc, implies external accelerations at the Solar Radius of slightly above $a_{0}$, e.g. \citet{Chae24a}. Thus, the expectation has been for disc open clusters to behave in the quasi-Newtonian regime, and not to present any of the deep-MOND phenomenology associated with the RAR and BTFR galactic scalings. However, the Galactic disc is not in fact a smooth double exponential matter distribution.

Aside from the presence of spiral structure, when looking at recent maps of matter distribution within $3\,$kpc of the Sun, typically using Gaia data (e.g. \citet{Poggio21}, \citet{Zari21}, \citet{Ge24}), a strong level of small-scale density fluctuations are apparent. The exact spiral structure of the Milky Way is still uncertain due to observational difficulties arising from our position within the Galaxy \citep{2014A&A...569A.125H, 2016SciA....2E0878X}. However recent measurements using Gaia have enabled mapping the spiral structure in the solar neighbourhood with a resolution of about 300 parsecs \citep{2024A&A...686A..70W}. This measurement shows that spiral arms cause a $\pm 20\%$ density perturbation. Combining the bulge and disc potential of \citet{1998MNRAS.294..429D} with the spiral potential of \citet{2002ApJS..142..261C} assuming a $\pm 20\%$ density perturbation shows that this alone is incapable of generating the necessary small-scale low-acceleration regions to explain the open cluster data arising from the deep-MOND regime. Smaller scale features such as HI clouds, molecular clouds, tidal streams and open clusters themselves also have their own gravitational fields which can counteract the galactic gravitational field if of sufficient mass and in the right proximity to a region of interest. 

To zeroth order, we can calculate from a Newtonian perspective that the 220 km/s of local Galactic rotation at a Galactocentric radius of $8.5\,kpc$ implies a total central mass of $10^{11} M_{\odot}$, the resulting gravitational force pointing towards the Galactic centre could then be balanced locally by a $10^{5} M_{\odot}$ molecular cloud at a distance of some 10 pc in the direction of the Galactic anti-centre. Thus, it is not impossible, indeed unavoidable, that the inhomogeneities of the small-scale Galactic disc matter distribution will lead to the existence of small scale pockets of overall gravitational potential minima where the external gravitational field will be much smaller than its mean value. Within these pockets, from a MOND perspective, one should expect a much reduced, or even zero EFE, and hence the existence of low internal acceleration systems showing deep-MOND phenomenology.

We propose that small low density gas clouds in the Galactic disc will generally be stable against their internal gravity, being mostly in the EFE dominated regime, but will become gravitationally unstable when entering any gravitational potential minima pockets, where they will find themselves close to the deep-MOND limit. These will then experience star formation and lead to small stellar open clusters with dynamical properties reflecting the BTFR and RAR scalings of galaxies. 
Indeed, it has long been held that molecular clouds and hence open clusters, form in spiral arm potential minima where material piles up as it passes through the arm on its orbit around the galaxy e.g. \citep{1987ApJ...314...10R}. On leaving these formation pockets small OCs enter again the EFE dominated regime and begin to dissolve into the field. This dissolution process will proceed through tidal effects, which are typically very sensitive to a tidal radius threshold. For example, when describing Roche lobe overflow in tight binary stars, tidal effects are ignored when describing the stellar structure within the Jacobi radius of the star being tidally stripped, while any material beyond this critical radius is lost to the tidal field of the accreting star. 

Similarly, the dynamical properties of the small OCs studied here, which closely follow the deep-MOND BTFR and RAR galactic scalings, have been determined exclusively within the tidal radii of each cluster. Stars in the tidal tails have been excluded from the analysis. The dynamics we describe therefore pertain only to the bound (or most nearly bound) inner regions that remain shielded from ongoing tidal stripping, as illustrated in Fig.~ ~\ref{FigScatterPlots}. For the more massive open clusters, the relevant distinction is not that their internal accelerations necessarily exceed $a_0$ in isolation, but that the total gravitational field experienced by the system, $g_{\rm tot} \simeq g_{\rm int} + g_{\rm ext}$, can lie closer to or above $a_0$, either currently or during formation. In such conditions MOND effects are expected to be strongly suppressed and the internal dynamics to approach the Newtonian virial expectation according to the forced Newtonian regime, with subsequent dissolution and tidal heating capable of driving some systems toward super-virial velocity dispersion values.

If the alignment of open clusters with the RAR and BTFR reflects low-acceleration MOND dynamics within the Galactic disc, this result has consequences that extend beyond cluster kinematics and bear directly on other probes of gravity and on the interpretation of the acceleration scale itself. For example OCs could afford a far stronger EFE test than \citet{2020ApJ...904...51C} if the external field could be quantified. This is because a much larger range of internal and external gravitational fields would be at play.

\subsection{Mapping the galactic gravitational field}
If open cluster velocity dispersions can be measured with improved precision and if some of these systems are confirmed to be in approximate dynamical equilibrium at least within their tidal radii, they could provide an additional method to map the Galactic gravitational field. In the MOND framework, their internal dynamics depend not only on their own baryonic mass distribution but also on the local external field through the external field effect. This introduces an environmental sensitivity that can, in principle, be inverted: given reliable cluster masses and velocity dispersions, the local external acceleration can be constrained. Open clusters would then serve as discrete, parsec-scale probes of the gravitational field within the disc. This approach is complementary to determinations based on the vertical motions of stars in the solar neighbourhood, such as those by \citet{2024A&A...686A..70W}, which infer the gravitational potential through large-scale kinematic modelling.

\subsection{Molecular clouds}
If the Galactic gravitational field varies on parsec scales and includes regions where the total external gravitational acceleration drops below $a_0$, molecular clouds can provide an independent test of this hypothesis. As the progenitors of open clusters and systems of comparable mass, they allow the same acceleration-based diagnostics to be applied in the gaseous phase. Modern Galactic cloud catalogues measure cloud sizes, line widths, and distances, permitting estimates of baryonic accelerations from the mass distribution and dynamical accelerations from CO line–derived velocity dispersions. Molecular clouds can therefore be placed in the same acceleration planes as stellar systems and directly compared with Newtonian and MOND expectations.

However, not all molecular clouds are suitable for such an analysis. Surveys that focus on dense cores primarily trace regions already affected by stellar feedback, pressure confinement, and local turbulence, where the dynamics are not dominated solely by self-gravity. At the opposite extreme, giant molecular cloud complexes identified at low angular resolution may blend multiple substructures, obscuring their true dynamical state. Meaningful tests therefore require clouds that are spatially resolved, trace the full extent of the system rather than only its densest parts, and are not strongly perturbed by active star formation. If such systems also exhibit alignment with the RAR or BTFR, this would strengthen the case that low-acceleration regions operate within the Milky Way disc and influence both gaseous and stellar self-gravitating systems.

\subsection{Wide binaries}
Similar to studies of molecular clouds, our current results could also be important to wide binary gravity tests. In these, the relative separations and velocities of wide binary stars are statistically compared to predictions under Newtonian and Milgromian gravity, \citet{Hernandez12}. Since wide binaries experience very low internal gravitational accelerations, they are expected to present a Newtonian gravitational anomaly.

Wide binary gravity tests explicitly exclude any objects being part or close to any stellar over-densities, open clusters or stellar associations, as part of requiring the wide binaries studied to be as isolated from any ambient dynamical perturbations as possible. Further, the clear presence of tidal tails in almost all OCs shows these systems are currently dissolving, under the MOND interpretation presented, we expect them to have been formed in local pockets of low EFE with dynamics which are still reflected in the kinematics of their stars inside their current tidal radii. The OCs' current positions do not necessarily reflect their initial formation sites a few hundred million years ago. Hence,  tracing the regions where OCs show more or less deep-MOND effects will probably not be very relevant to wide binary gravity tests.

However, if the Galactic gravitational field is spatially variable and includes regions where it falls below $a_0$, as our current results suggest, then MOND predicts wide binaries to exhibit a substantial range of external field effect strengths. This suggests significant variations in wide binary gravity results might be expected, particularly once observational accuracy increases to the point where these tests can extend beyond the current $\sim 150\,$pc volume probed, out to kpc-scales where a greater range of total gravitational acceleration will be probed.

\subsection{Acceleration scale convergence}

\begin{table}
\caption{Determinations of Milgrom's constant $a_0$
($10^{-10}\,\mathrm{m\,s^{-2}}$).}
\label{a0table}
\centering
\begin{tabular}{lcc}
\hline\hline
Reference & $N^{a}$ & $a_0$ \\
\hline
\multicolumn{3}{l}{\textit{Baryonic Tully--Fisher relation}} \\
\citet{1991MNRAS.249..523B} & 10  & $1.21 \pm 0.24$ \\
\citet{2009AJ....138..392S}$^{e}$ & 28  & $1.18^{d}$ \\
\citet{2009AA...505..577T}$^{e}$ & 34  & $1.30^{d}$ \\
\citet{2011PhRvL.106l1303M}$^{e}$ & 47  & $1.24 \pm 0.14$ \\
\citet{2016ApJ...816L..14L} & 118 & $1.29 \pm 0.06^{f}$ \\
\hline
\multicolumn{3}{l}{\textit{Central surface density relation}} \\
\citet{2009MNRAS.397.1169D} & $\sim 10^{3}$ & $1.3^{g}$ \\
\citet{2016ApJ...827L..19L}$^{h}$ & 135 & $1.27 \pm 0.05^{i}$ \\
 &  & $1.27 \pm 0.05^{j}$ \\
\hline
\multicolumn{3}{l}{\textit{Radial acceleration relation}} \\
\citet{2015MNRAS.446..330W} & 74  & $0.94 \pm 0.03^{i}$ \\
 &  & $1.21 \pm 0.03^{k}$ \\
\citet{2016PhRvL.117t1101M} & 153 & $1.20 \pm 0.02^{l} \pm 0.24^{m}$ \\
\citet{2017ApJ...836..152L} &  &  \\
\textbf{Open clusters} $\mathbf{N_\star \leq 250}$ & $\mathbf{3251^{n}}$ & $\mathbf{\sim1.4 \pm 0.54dex}$ \\
\textbf{Open clusters fiducial} & $\mathbf{2423^{n}}$ & $\mathbf{\sim1.2 \pm 0.52dex}$ \\
\hline
\end{tabular}

\vspace{0.2cm}
\footnotesize
$^{a}$ Number of galaxies. 
$^{d}$ From \citet{2012AJ....143...40M}. 
$^{e}$ Gas-rich galaxies only. 
$^{f}$ Corrected for non-point-mass potential. 
$^{g}$ Based on assumed projected dark matter density. 
$^{h}$ Least-squares fit to \citet{2016ApJ...827L..19L}. 
$^{i}$ $\nu(y)$ with $n=1$. 
$^{j}$ $\nu(y)=(1-e^{-\sqrt{y}})^{-1}$. 
$^{k}$ $\nu(y)$ with $n=2$. 
$^{l}$ Random error. 
$^{m}$ Systematic error.
$^{n}$ Note that individual galaxies are usually measured at 20-50 points along the disc so the total number of data points is actually similar between galaxies and open clusters.
\end{table}

A further consideration concerns the convergence of the acceleration scale $a_0$ across independent empirical domains. As emphasized in philosophical analyses of theory assessment such as \citet{Merritt}, a parameter that repeatedly emerges from distinct phenomena carries greater evidential weight than one inferred multiple times from essentially the same class of data. Within the Milgromian framework, $a_0$ has been recovered from several independent scaling relations, including the baryonic Tully–Fisher relation, the radial acceleration relation, and the central surface density relation. Table \ref{a0table} reproduces and updates Table 8.1 of \citet{Merritt}, which is based on different galaxy samples and methodologies and yet cluster around a common value.

The consistency of the best-fit $a_0$ obtained from open clusters extends this convergence to a qualitatively different dynamical regime. Open clusters are parsec-scale stellar systems embedded within the Galactic disc and characterised by velocity dispersions rather than rotation curves. Agreement in the inferred acceleration scale across these distinct systems and observables strengthens the interpretation of $a_0$ as a genuine dynamical constant rather than a parameter adjusted to fit a particular dataset.

\section{Conclusions}
We have investigated whether open clusters in the Milky Way follow the radial acceleration relation (RAR) and the baryonic Tully--Fisher relation (BTFR), scaling laws that were originally established for galaxies on kiloparsec scales. Using 3618 Gaia DR3 open clusters from the HR134 catalogue, we computed baryonic and observed accelerations in a manner consistent with Lelli et al.\ (2017) and placed these systems in the acceleration and mass--velocity planes.

Approximately 90\% of open clusters (those with $N_\star \leq 250$) lie close to the RAR, albeit with substantial scatter. A first fit of the RAR to a high-quality subsample yields a best-fit acceleration scale $g_\dagger \approx 1.2 \times 10^{-10}\,\mathrm{m\,s^{-2}}$ with an uncertainty of about 0.5 dex. This value is consistent with canonical determinations of $a_0$, though far less precise. More massive clusters tend to approach the Newtonian virial expectation, suggesting a transition regime that is qualitatively compatible with an external field contribution. No correlations are present between RAR residuals and galactocentric radius, distance to the Galactic disc midplane, age, ellipticity, tidal tail fraction or cleanliness of the CMD.

Alternative explanations for the elevated velocity dispersions were examined. Interloper contamination and unresolved binaries would increase the inferred stellar velocity dispersions of the OCs explored, but explaining the observed trends would require fine-tuning these effects in a contrived manner. Reclassifying the majority of systems as unbound associations does not naturally explain their concentration along the RAR, since dissolving systems could in principle occupy a broad region above the Newtonian virial line. 

If the alignment of most open clusters with the RAR and BTFR reflects MOND low-acceleration dynamics, this implies that parsec-scale systems within the Milky Way disc experience gravitational environments that are not adequately described by a spatially smooth field of order $a_0$. This would indicate that the effective external field varies on small scales, reducing the suppressive impact of the external field effect in some regions. Open clusters would then serve as discrete probes of the Galactic gravitational field on parsec scales.

The consistency of the inferred acceleration scale with values obtained from galaxies extends the convergence of $a_0$ to a distinct dynamical regime characterised by pressure-supported stellar systems embedded within a galactic disc. The quantitative consistency we find between OC properties and the galactic BTFR and RAR scalings motivates improved modelling of cluster masses, velocity dispersions, and the local external field. Higher-resolution studies of the Galactic potential and independent analyses of molecular clouds and wide binaries will be essential to clarify whether the observed behaviour represents a small-scale manifestation of the same acceleration scale that governs galactic dynamics.

\section*{Acknowledgements}
Gratitude goes to Anthony Brown for comments that helped to understand the systematics of Gaia DR3. X.H. acknowledges financial assistance from SECIHTI SNII and UNAM DGAPA PAPIIT grant IN-102624.

\section*{Data availability}

All data used will be shared upon reasonable request to the authors.

\bibliographystyle{mnras}
\bibliography{ocs}

\bsp
\label{lastpage}
\end{document}